\documentclass[12pt]{article}
\usepackage{setspace}
\usepackage{amsmath,amssymb}
\usepackage{color}
\usepackage{hyperref}
\usepackage{array}
\usepackage{booktabs}
\numberwithin{equation}{section}
\usepackage{graphicx}
\usepackage{braket}

\usepackage{soul}

\newcommand{\beq}{\begin{equation}}
\newcommand{\eeq}[1]{\label{#1}\end{equation}}
\newcommand{\bea}{\begin{eqnarray}}
\newcommand{\eea}[1]{\label{#1}\end{eqnarray}}

\def\mL{\mathcal{L}}

\def\mM{\mathcal{M}}
\def\mO{\mathcal{O}}

\def\mM{\mathcal{M}}

\def\mI{\mathcal{I}}

\def\zb{\bar{z}}

\def\qb{q}
\renewcommand{\hbar}{} %\hbar = 1 units

\def\mb{\bar{m}}

\def\rb{\bar{r}}

\def\pa{\partial}

\def\g5{\gamma_5}

\def\Qt{\tilde{Q}}

\def\b[#1]{\bold{#1}}
\def\bb[#1]{\overline{\bold{#1}}}
\def\bs[#1,#2]{\bold{#1}_{#2}}
\def\bbs[#1,#2]{\overline{\bold{#1}}_{#2}}

\def\s2{\sigma_2}

\def\gammaflat{ \gamma_{z\zb}}

\def\Tsoft{T_{\text{soft}}}
\def\Thard{T_{\text{hard}}}

\def\ketd[#1]{\ket{#1}_{\text{dressed}}}
\def\brad[#1]{\bra{#1}_{\text{dressed}}}
\def\ketas[#1]{\ket{#1}_{\text{Asymptotic}}}
\def\braas[#1]{\bra{#1}_{\text{Asymptotic}}}

\def\eval{\, \! \big|}

\def\gammaflat{ \gamma_{z\zb}}

\def\Tsoft{T_{\text{soft}}}

\def\bF{\bold F}
\def\bp{\bold p}
\def\bA{\bold A}
\def\bB{\bold B}
\def\bD{\boldsymbol{\nabla} }
\def\br{\bold r}
\def\bE{\bold E}
\def\bA{\bold A}
\def\bv{\bold v}
\def\dd{\hat d}
\def\del{\hat \delta}
\def\eq{\begin{equation}}
\def\eqe{\end{equation}}
\def\eqa{\begin{eqnarray}}
\def\eqae{\end{eqnarray}}

\begin{document}
	\setlength{\topmargin}{-1cm} 
	\setlength{\oddsidemargin}{-0.25cm}
	\setlength{\evensidemargin}{0cm}

\begin{titlepage}
\hfill NCTS-TH/1908
\begin{center}

\vskip 2 cm

{\LARGE \bf  The Double Copy of Electric-Magnetic Duality}

\vskip 2 cm

{Yu-tin Huang$^{1,2}$, Uri Kol$^3$ and Donal O'Connell$^4$}

\vskip .75 cm

{	\em
	$^1$
	\em Department of Physics and Astronomy, National Taiwan University, Taipei 10617, Taiwan\\
	$^2$
	Physics Division, National Center for Theoretical Sciences,
	National Tsing-Hua University, No.101, Section 2, Kuang-Fu Road, Hsinchu, Taiwan\\
	$^3$
	Center for Cosmology and Particle Physics,
	Department of Physics, New York University,
	726 Broadway, New York, NY 10003, USA\\
	$^4$
	Higgs Centre for Theoretical Physics, School of Physics and Astronomy,
    The University of Edinburgh, Edinburgh EH9 3JZ, Scotland, UK
}
	\vspace{12pt}

\end{center}

\vskip 1.25 cm

\begin{abstract}
\noindent
%In this paper, we argue that the complex transformation relating the Schwarzschild metric to that of Taub-NUT, introduced by Talbot, is in fact an electric-magnetic duality transformation. We show that at null infinity, the complex transformation can be mapped to a BMS supertranslation, which rotates the supertranslation and the dual (magnetic) supertranslation charge. This can also be seen from the cubic coupling between the classical source and its background, which for Taub-NUT is given by a complex phase rotation acting on gravitational minimal couplings. This is the same phase rotation generating dyons from photon minimal couplings, manifesting the double copy relation between the two solutions.
%
We argue that the complex transformation relating the Schwarzschild to the Taub-NUT metric, introduced by Talbot, is in fact an electric-magnetic duality transformation.
We show that at null infinity, the complex transformation is equivalent to a complexified BMS supertranslation, which rotates the supertranslation and the dual (magnetic) supertranslation charges. This can also be seen from the cubic coupling between the classical source and its background, which for Taub-NUT is given by a complex phase rotation acting on gravitational minimal couplings. The same phase rotation generates dyons from electrons at the level of minimally coupled amplitudes, manifesting the double copy relation between the two solutions.

\end{abstract}
\end{titlepage}
\newpage

%%%%%%%%%%%%%%%%%%%%%%%%%%%%%%%%%%%%%%%%%%%%%%%%%%%%%%%%%%%%%%%%
%%%%%%%%%%%%%%%%%%%%%%%%%%%%%%%%%%%%%%%%%%%%%%%%%%%%%%%%%%%%%%%%
\section{Introduction}
%%%%%%%%%%%%%%%%%%%%%%%%%%%%%%%%%%%%%%%%%%%%%%%%%%%%%%%%%%%%%%%%
%%%%%%%%%%%%%%%%%%%%%%%%%%%%%%%%%%%%%%%%%%%%%%%%%%%%%%%%%%%%%%%%

A solution-generating technique known in the literature as the \emph{complex coordinate transformation} provides a set of maps between different solutions of General Relativity.
Newman and Janis \cite{Newman:1965tw} were the first to observe that one can re-derive the Kerr metric by complexifying the Schwarzschild solution in null polar coordinates and then performing a shift. The construction was later extended to a class of solutions including the Kerr-Newman black hole and the Taub-NUT metric \cite{Talbot:1969bpa}. However, while the technical algorithm that defines this set of maps has been known for many decades, the underlying physical mechanism behind it remains obscure.

Recently, progress in understanding the map from the Schwarzschild solution to the Kerr metric was made in \cite{Arkani-Hamed:2019ymq}. The authors of \cite{Arkani-Hamed:2019ymq} argued that this complex shift originates from the exponentiation of spin operators in the minimally coupled three-point amplitude of a spinning particles and a \emph{graviton}.
This was demonstrated by computing the impulse imparted to a test particle by the Kerr black hole. In particular, it was shown that the exponentiation induces a complex shift to the impact parameter, transforming the impulse of a Schwarzschild black hole to that of Kerr. This complex shift was also shown to be related by the double copy\footnote{See~\cite{Bern:2019prr} for a comprehensive review of the double copy.} structure, which identifies solutions of the Einstein equations as the square of Yang-Mills gauge theory solutions, to a similar exponentiation of spin operators in the minimal coupling of spinning particles to a \emph{photon}. More generally, the Newman-Janis link between the Reissner-Nordstr\"om and Kerr-Newman black holes was also connected to minimal coupling and spinning particles~\cite{Moynihan:2019bor}.

\begin{figure}[]
	\begin{center}
		\includegraphics[scale=1.5]{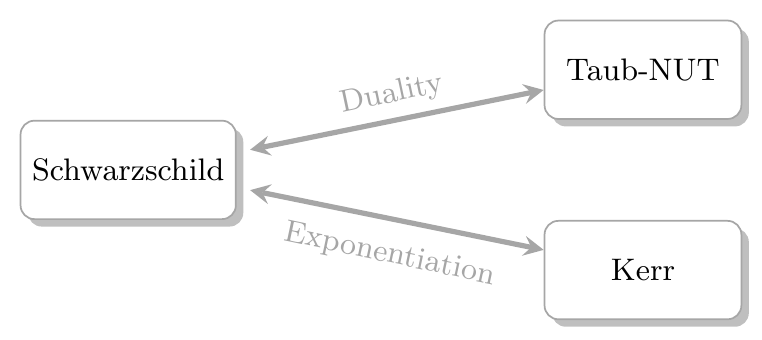}
	\end{center}
	\caption{
		The maps from the Schwarzschild solution to the Taub-NUT and to Kerr metrics.
		The underlying physical mechanism behind these two maps is different. In the first case the map is a duality operation while in the later case it results from exponentiation of spin operators \cite{Arkani-Hamed:2019ymq}.
		It would be interesting to extend this diagram and understand the full set of maps between solution in the Plebanski-Demianski family~\cite{Debever:1971,Plebanski:1976gy}.
	}
	\label{maps}
\end{figure}

There is another exponentiation that one can apply to minimal coupling: a pure phase rotation. For the case of coupling to photons, the rotation is an electric-magnetic duality transformation that maps an electrically charged particle to a dyon. 
The gravitational counterpart of this phase shift can be argued to map the Schwarzschild solution to the Taub-NUT metric.
Indeed, as was shown by one of the authors~\cite{Luna:2015paa}, the Taub-NUT metric admits a double copy structure whose ``square-root" is precisely the electromagnetic dyon.
This suggests that the exponentiated phase shift once again is related to some complex coordinate transformation which can be identified as some form of a gravitational electric-magnetic  duality. We depict this picture in figure \ref{maps}.

That scattering amplitudes can be used to understand the dynamics of magnetically charged particles was recently emphasised by Caron-Huot and Zahraee~\cite{Caron-Huot:2018ape}. It is worth emphasising that in our application, the perturbative expansion parameter is not the product of charges, which is quantised, but rather the ratio of this product to a classical angular momentum. In other words, we expand in a small scattering angle. Thus, the usual obstruction to amplitudes for magnetically charged particles is not present. Another interesting fact is that, in the presence of a magnetic charge, the scattering amplitudes themselves are not gauge invariant~\cite{Caron-Huot:2018ape}. However, the observable which we will construct from the amplitude is fully gauge invariant as it must be.

The complex coordinate transformation that maps the Schwarzschild and Taub-NUT solutions into each other was studied by Talbot \cite{Talbot:1969bpa}, who generalized the Newman-Janis algorithm and provided a set of mappings between solutions obeying a generalized Kerr-Schild form.
This includes a map between the Schwarzschild metric, which is given by
 \begin{equation}
ds^2 = -f_\text{Sch}(r) dt^2 +\frac{1}{f_{\text{Sch}}} dr^2+r^2 (d \theta ^2 +\sin ^2 \theta d \varphi^2 ), \qquad f_{\text{Sch}}(r) = 1- \frac{2m}{r},
\end{equation}
and the Taub-NUT metric
\begin{equation}\label{TaubNUTMetric}
\begin{aligned}
ds^2 &= 
-f_{\text{NUT}} \left(dt +2 \ell \cos \theta d \varphi\right)^2
+\frac{1}{f_{\text{NUT}}} dr^2
+\left(r^2 + \ell^2 \right)\left(d\theta^2 +\sin ^2 \theta d\varphi ^2\right), \\
f_{\text{NUT}} (r) &= \frac{r^2 -2mr -\ell^2}{r^2 +\ell^2},
\end{aligned}
\end{equation}
where $m$ is the ADM mass parameter and $\ell$ is the NUT charge. 
The map between the two solutions is given by the following complex coordinate transformation
 \begin{equation}\label{mapCoord}
u \rightarrow u   + 2 i \ell \log \sin \theta, \qquad
r \rightarrow r  - i \ell ,
\end{equation}
 accompanied by a map of the parameters
\begin{equation}\label{globalTran}
m \rightarrow m - i \ell.
\end{equation}
For more information about the complex coordinate transformation we refer the reader to \cite{Keane:2014sta}-\cite{Erbin:2016lzq}.

Our main goal in this paper is to understand the physical origin of the complex coordinate transformation \eqref{mapCoord} and relate it to the phase rotation of the three-point amplitude between massive particles and gravitons.
First, we show that the complex coordinate transformation \eqref{mapCoord} is in fact a BMS supertranslation transformation with a complex parameter. The complex parameter renders the transformation a map rather than a symmetry.
We then show that under this map the supertranslation charge $T(f)$ of the Schwarzschild black hole transforms into a complex linear combination of standard and dual supertranslations
\begin{equation}\label{duality}
	T(f) \rightarrow T(f) - i \mM (f),
\end{equation}
where $\mM  (f)$ is the dual supertranslation charge of the Taub-NUT metric \cite{Kol:2019nkc}.
In particular, the global parts of the charges, which are given by $T_{\text{global}}=m$ and $\mM_{\text{global}}=\ell$, transform as \eqref{globalTran}.
This result provides a physical explanation for the map between the Schwarzschild and Taub-NUT solutions as a duality transformation.

Finally, we demonstrate that the impulse imparted to a test particle is indeed reproduced by the phase rotation of the minimally coupled amplitude. We first demonstrate it in the electromagnetic case for a probe particle moving under the influence of the background dyon fields.
In the gravitational case, the double copy structure of the three point amplitude implies that the phase shift is doubled and accordingly we show that the resulting minimal coupling indeed reproduces the impulse of a test particle moving in the Taub-NUT background, as computed from the classical geodesic equations. While the double copy structure is manifest in the amplitude description, it is less apparent from the classical equations of motion. We use our analysis to demonstrate how the double copy structure relates the geodesic equations to the electromagnetic Lorentz force experienced by the probe particle.

The Taub-NUT solution is known in the literature as the gravitational analogue of an electromagnetic dyon, where the ADM mass and NUT charge correspond to the electric and magnetic charges of the dyon, respectively. In this paper we further elaborate on the analysis of \cite{Luna:2015paa} and show that this correspondence is not merely just an analogy but rather a map that is dictated by the double copy structure. The double copy maps the large electric and magnetic $U(1)$ charges in QED into the standard and dual supertranslation charges in gravity, respectively. As a consequence, the duality \eqref{duality} is therefore mapped into the electric-magnetic duality in Yang-Mills theory.

This paper is organized as follows.
In section \ref{BMS} we show that the complex coordinate transformation \eqref{mapCoord} that maps Schwarzschild to Taub-NUT is a complex BMS supertranslation transformation. In section \ref{dualitySection} we show that this complex transformation rotates the standard and dual supertranslation charges into each other and therefore we interpret the transformation as a duality operation. In section \ref{impulseSection} we compute the impulse on a test particle in the presence of a dyon, in the electromagnetic case, and due to its motion on the Taub-NUT background, in the gravitational case. We perform the computation in two ways: using the three-point amplitude, and using the classical equations of motion. The double copy structure is manifest from the amplitude perspective and we use our results to show how it arises in the classical equations of motion. We conclude in section \ref{ConclusionSection} with a discussion followed by appendices reviewing the standard complex coordinates transformation algorithm and some technical computational details.

\textbf{Note added:} While this manuscript was in preparation, reference~\cite{Alawadhi:2019urr} appeared which has overlap with this paper.

\section{Complex BMS Supertranslations}\label{BMS}

In this section we show that the transformation \eqref{mapCoord} is in fact an imaginary BMS supertranslation transformation to leading order in the asymptotic expansion.
We first review some standard notation. It will be convenient for us to use complex coordinates $z$ and $\zb$ on the sphere, related to the standard angle variables by
\begin{equation}
z= \tan \left(\frac{\theta}{2}\right) e^{i \varphi}.
\end{equation}
Indices on the sphere are raised and lowered with the standard round metric $\gammaflat$ on the unit sphere $S^2$, given in these coordinates by
\begin{equation}
\gammaflat = \frac{2}{(1+z \zb)^2} \,.
\end{equation}
We denote the covariant derivatives on the sphere by $D_z,D_{\zb}$. It is helpful to note that the only two non-vanishing Christoffel symbols on the sphere are
\begin{equation}
\begin{aligned}
\Gamma^z_{zz} & = \gamma^{z\zb} \pa_z \gammaflat = - \gammaflat \pa_z \gamma^{z\zb} = - \frac{2\zb}{1+z \zb}, \\
\Gamma^{\zb}_{\zb\zb} & = \gamma^{z\zb} \pa_{\zb} \gammaflat = - \gammaflat \pa_{\zb} \gamma^{z\zb} = - \frac{2 z}{1+z \zb}.
\end{aligned}
\end{equation}
In particular, these results imply that the action of the covariant derivative on a vector in the sphere's directions is
\begin{equation}
\begin{aligned}
D_z V_z &= \pa_z V_z - \Gamma ^z _{zz} V_z, \\
D_z V_{\zb} &= \pa_z V_{\zb},
\end{aligned}
\end{equation}
and similarly for $D_{\zb}$. The generalization to higher rank tensors is straightforward. In particular, it is useful to note that the action of the double covariant derivative on a scalar function $S(z,\zb)$ is given by
\begin{equation}
D_z^2 S = \pa_z^2 S -  \Gamma ^z _{zz} \pa_z S = \gammaflat \pa_z \left(\gamma^{z\zb} \pa_z S\right)=\gammaflat \pa_z \pa^{\zb} S.
\end{equation}
Finally, let us also mention that the only non-vanishing components of the Ricci and Riemann tensors are
\begin{equation}
\begin{aligned}
R_{z\zb} &= \gammaflat, \\
R_{z \zb \zb z} &= R_{\zb z z \zb} = - R_{z \zb z \zb} =- R_{\zb z \zb z } = \gammaflat ^2 .
\end{aligned}
\end{equation}

The expansion of asymptotically flat metrics around future null infinity $\mathcal{I}^+$ in the Bondi gauge is given by
\begin{equation}\label{futureMetric}
\begin{aligned}
ds^2 &= -du^2 -2dudr+2r^2 \gammaflat dz d\zb \\
&+
\frac{2m_B}{r} du^2 +r C_{zz} dz^2 +r C_{\zb\zb} d\zb ^2 -2U_z du dz -2 U_{\zb} du d\zb 
\\
&+ \dots ,
\end{aligned}
\end{equation}
where $u$ is the retarded time, 
\begin{equation}
U_z = -\frac{1}{2} D^z C_{zz} \,,
\end{equation}
and the dots indicate subleading terms in the expansion around $r= \infty$.
The symbol $m_B$ denotes the Bondi mass aspect. The Bondi news, given by
\begin{equation}
N_{zz} = \pa_u C_{zz},
\end{equation}
characterizes gravitational radiation. At spatial infinity the radiative data is given by
\begin{equation}\label{boundaryGraviton}
C_{zz} \eval _{\mI^+_{-}} = D_z^2 C (z , \zb),
\end{equation}
where $C(z,\zb)$ is a complex scalar function called the \emph{boundary graviton} \cite{Kol:2019nkc}.

The metric \eqref{futureMetric} is invariant under the BMS group. Here we will focus on one component of this group, known as the \emph{BMS supertranslations}, whose generator is given by
\begin{equation}\label{superT}
\begin{aligned}
T(f) &= \frac{1}{4\pi G}
\int_{\mI^+_-} d^2 z \gammaflat f(z,\zb) m_B.
\end{aligned}
\end{equation}
Here $f(z,\zb)$ is a \emph{real} transformation parameter.
The action of supertranslations on the metric is described by the vector field
\begin{equation}\label{coordTrans}
\xi_f = f \pa_u-\frac{1}{r} \left(D^z f \pa_z +D^{\zb} f \pa_{\zb} \right)
+D^z D_z f \pa_r
+\mO \left( r^{-2} \right),
\qquad
f= f(z,\zb).
\end{equation}
In particular, it implies that the radiative data and the boundary graviton transform as
\begin{equation}\label{metricTrans}
\begin{aligned}
\mL_f C_{zz} &=f \pa_u C_{zz}-2D_z^2 f , \\
\mL_f C_{\zb \zb } &=f \pa_u C_{\zb \zb}-2D_{\zb}^2 f , \\
\mL_f C&=-2 f  .
\end{aligned}
\end{equation}
Note that supertranslation transformations are symmetries only when $f(z,\zb)$ is a real function.

In \cite{Kol:2019nkc} it was shown that the Taub-NUT metric, equation \eqref{TaubNUTMetric}, can be brought to the Bondi form \eqref{futureMetric} with
\begin{equation}
C_{zz} =  i \ell \gammaflat \frac{1+|z|^4}{z^2}
\end{equation}
everywhere except at the location of the string singularity. This implies that the boundary graviton is given by
\begin{equation}\label{ImCTN}
C (z,\zb)= 4 i \ell     \log  \frac{1+|z|^2}{2|z|}  .
\end{equation}
(the normalization of the argument inside the logarithm is completely arbitrary; we choose the factor 2 for convenience).
This result implies that, to leading order in the asymptotic expansion, the Schwarzschild solution can be mapped into the Taub-NUT metric using an imaginary BMS supertranslation transformation with a parameter
\begin{equation}\label{tranPara}
f (z,\zb)=  - 2 i \ell     \log  \frac{1+|z|^2}{2|z|}  = 2i \ell \log  \sin \theta.
\end{equation}
With this imaginary supertranslation parameter the transformation describes a map rather than a symmetry. The coordinate transformation \eqref{coordTrans} now takes the form
\begin{equation}
\xi_{\ell} = (2i \ell \log  \sin \theta) \pa_u
-(i \ell  )  \pa_r
-\frac{4i \ell }{r} 
\cot \theta d \theta
+\mO \left( r^{-2} \right).
\end{equation}
Up to the order $\mO(r^{-1})$ term, which is needed only in order to preserve the Bondi gauge, this transformation precisely coincides with \eqref{mapCoord}.

\section{Duality Transformation}\label{dualitySection}

Asymptotically flat spacetimes are invariant under the BMS group which contains supertranslations and Lorentz transformations.
Recently it was found that spacetimes which are \emph{locally} asymptotically flat possess a larger symmetry group that contains, in addition, a \emph{dual supertranslation} symmetry \cite{Kol:2019nkc,Godazgar:2018qpq,Godazgar:2018dvh}. The new symmetry defines a conserved dual supertranslation charge
\begin{equation}\label{dualSuperT}
\begin{aligned}
\mM(\varepsilon) =
\frac{i}{16 \pi G} \int_{\mI^+_-}  d^2 z\, \varepsilon(z,\zb) \gamma^{z\zb} \left(D^2_{\zb}C_{zz}-D^2_z C_{\zb\zb}\right)
\end{aligned}
\end{equation}
which is akin to the large magnetic $U(1)$ charge in Quantum Electrodynamics (QED).
By the same analogy, the supertranslation charge is akin to the large electric $U(1)$ charge in QED.
In the following we will relate the complex coordinate transformation discussed in the previous sections to a duality operation on these asymptotic charges.

First of all, we use standard techniques to decompose the supertranslation charge \eqref{superT} into soft and hard parts
\begin{equation}
T(\varepsilon) = \Tsoft (\varepsilon) + \Thard (\varepsilon),
\end{equation}
where the soft part of the charge receives contributions from zero-energy graviton modes only. 
This is done using the $uu$ component of the Einstein equations $G_{uu} = 8\pi T^M_{uu}$
\begin{equation}\label{uuEquation}
\begin{aligned}
\pa_u m_B &= \frac{1}{4} \left(D_z^2 N^{zz}+D_{\zb}^2 N^{\zb\zb}\right) - T_{uu} ,\\
T_{uu} &=  4\pi G \lim_{r \rightarrow \infty } \left(r^2 T^M_{uu}\right) + \frac{1}{4} N_{zz}N^{zz}.
\end{aligned}
\end{equation}
where $T^M_{uu}$ is the $uu$ component of the stress-energy tensor.
The supertranslation charge \eqref{superT} is given by an integral over the two sphere at spatial infinity. Using the Einstein equation \eqref{uuEquation} we can invert it into a three dimensional integral that includes the null coordinate $u$ as well. The resulting hard and soft contributions to the supertranslation charge are then given by
\begin{equation}
\begin{aligned}
\Thard (\varepsilon) 
&= \frac{1}{4\pi G} \int _{\mI^+} du d^2 z \,\varepsilon(z,\zb) \gammaflat T_{uu},
\\
\Tsoft (\varepsilon)  &=
-\frac{1}{16 \pi G} \int_{\mI^+} du d^2 z\, \varepsilon(z,\zb) \gamma^{z\zb} \left(D^2_{\zb}N_{zz}+D^2_z N_{\zb\zb}\right).
\end{aligned}
\end{equation}
Using this decomposition one can now show that under the transformation \eqref{coordTrans}-\eqref{metricTrans} the supertranslation charge transforms as
\begin{equation}
T(\varepsilon) \rightarrow T(\varepsilon) + \frac{1}{4\pi G} \int_{\mI^+_-} d^2 z \, \varepsilon(z,\zb) \gamma^{z\zb} D^2_z D^2_{\zb} f(z,\zb),
\end{equation}
with $f(z,\zb)$ given by \eqref{tranPara}. Since $D^2_zD^2_{\zb}f=- i \ell \gammaflat ^2 $ we then find
\begin{equation}
T(\varepsilon) \rightarrow T(\varepsilon) - i \frac{ \ell}{4\pi G} \int_{\mI^+_-} d^2 z \, \gammaflat  \varepsilon(z,\zb) .
\end{equation}
It is now evident that under the coordinate transformation \eqref{coordTrans} with a complex parameter \eqref{tranPara} the supertranslation charge receives a complex contribution. This complex contribution precisely coincides with the dual supertranslation charge of the Taub-NUT solution, as was found in \cite{Kol:2019nkc}. Therefore the supertranslation charge transforms as
\begin{equation}
T(\varepsilon) \rightarrow T(\varepsilon) - i \mM_{\text{NUT}} (\varepsilon).
\end{equation}
under the complex coordinate transformation.
The parameter $\ell$ is known as the NUT charge or equivalently as the magnetic mass aspect.
The global parts of the charges, described by a constant parameter $\varepsilon$, are given by $T_{\text{global}}=\frac{m}{G}$ and $\mM_{\text{global} }=\frac{\ell}{G}$, which therefore transform as
\begin{equation}
m  \rightarrow m - i \ell .
\end{equation}
We therefore see that the complex coordinate transformation \eqref{mapCoord} in fact generates a duality transformation that mixes the standard and dual supertranslation charges, or equivalently the mass and the magnetic mass aspects! This provides a physical explanation for why the complex coordinates transformation defines a map between different solutions.
We can now define a complex charge whose real and imaginary parts are given by the standard and dual supertranslation charges
\begin{equation}
Q (\varepsilon) \equiv T(\varepsilon) - i \mM (\varepsilon) \equiv C (\varepsilon)  e^{i \theta(\varepsilon)}.
\end{equation}
Duality transformations therefore simply correspond to rotations of the phase $\theta$.

Let us comment that from this perspective the complex transformation function \eqref{tranPara} is not unique. Different complex transformation functions will give rise to different NUT-type solutions characterized by their dual supertranslation charges (in addition, of course, to their supertranslation charges). While the function \eqref{tranPara} describes an infinite gravito-magnetic string, different functions will give rise to conical singularities of different types. For example, the function
\begin{equation}
f (z,\zb)=  - 2 i \ell     \log  (1+|z|^2)  = 4i \ell \log  \cos \frac{\theta}{2}
\end{equation}
will describe a semi-infinite string pointing in the direction $\theta=\pi$ (corresponding to $z=\infty$). More generally we can have a structure of multiple isolated singularities on the celestial sphere corresponding to several semi-infinite strings pointing in different directions.

We now want to study the effect of the duality transformation on the Coulomb part of the gravitational field. For this purpose we will use the Newman-Penrose (NP) formalism \cite{Newman:1961qr,Newman:1968uj}. Newman and Penrose grouped all the curvature invariants into five complex scalars
\begin{equation}
\Psi_n, \qquad n=0,\dots,4.
\end{equation}
The five complex scalars decay at infinity as $\Psi_n \sim r^{-5+n}$ and they classify spacetime into 5 different zones according to their decay rate. These zones are analogous to the near, intermediate and far zones in electrodynamics, except that in gravity there are five zones. The Coulomb component of the gravitational field is described by the complex scalar $\Psi_2$, in the same way that in electrodynamics the Coulomb potential is dominant in the intermediate zone, while the rest of the scalars describe radiative modes. For the Schwarzschild metric the Coulomb potential takes the form
\begin{equation}
\Psi_2 = \frac{m}{r^3}.
\end{equation}
Now under the duality transformation the Coulomb potential transforms into
\begin{equation}
\Psi_2 = \frac{m-i \ell}{\left(r-i\ell \right)^3},
\end{equation}
which is precisely the value of $\Psi_2$ in the Taub-NUT background. 

We have therefore shown that the complex coordinate transformation \eqref{mapCoord}  generates a duality transformation that maps the charges and the Coulomb potential of the Schwarzschild solution into those of the Taub-NUT solution.

%%%%%%%%%%%%%%%%%%%%%%%%%%%%%%%%%%%%%%%%%%%%%%%%%%%%%%%%%%%%%%%%
%%%%%%%%%%%%%%%%%%%%%%%%%%%%%%%%%%%%%%%%%%%%%%%%%%%%%%%%%%%%%%%%
\section{The on-shell phase rotation}\label{impulseSection}
\begin{figure}
\begin{center}
\includegraphics[scale=0.6]{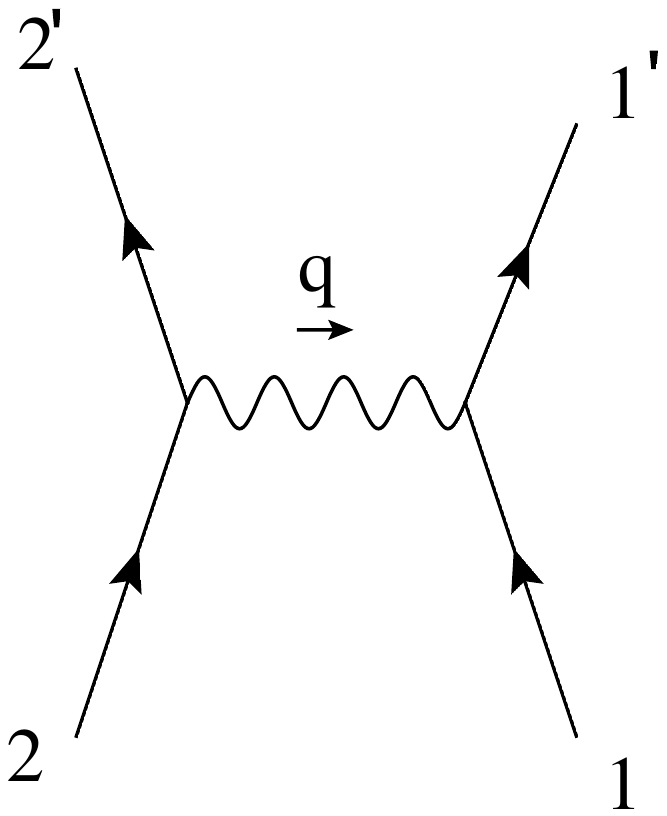}
\caption{We consider the gravitational and electromagnetic impulse imparted on a probe particle 1 via the source particle $2$, at leading order in coupling. The effect is captured by the one graviton or photon exchange respectively.}
\label{fig1}
\end{center}
\end{figure}

%%%%%%%%%%%%%%%%%%%%%%%%%%%%%%%%%%%%%%%%%%%%%
In the previous sections we have seen that the complex shift generated by BMS supertranslations transforms the Schwarzschild solution into the Taub-NUT solution. Furthermore, when viewed from the point of view of conserved charges, it induces a phase rotation between the ``electric" and the ``magnetic" BMS charge. In the following we will demonstrate this phenomenon directly using on-shell observables. In particular, we will compute the impulse on a probe in the classical background. This can be captured~\cite{Kosower:2018adc} from the $2\rightarrow 2$ scattering of the probe off a very heavy, static particle sourcing the classical background. At leading order, the classical impulse is
\begin{align}\label{eq:impulseGeneral}
\Delta p_1^\mu =\frac{1}{ 4m_1m_2} \int \dd^4 &\qb\;\del(\qb\cdot u_1)\del(\qb\cdot u_2)e^{{-}i\qb\cdot b} \, i\qb^\mu \, 
M_4\left(1,2\rightarrow 1',2'\right)|_{\qb^2\rightarrow 0} \,,
\end{align} 
where particle $1$ is the probe while particle 2 is the classical source, and $u_1,u_2$ are their proper velocities. The momentum transfer between the two particles is labeled by $q$.
The kinematic setup is shown in figure \ref{fig1}. As one can see, we will be interested in the $\qb^2\rightarrow 0$ limit of the $2\rightarrow 2$ scattering since we are interested in long range effects. 

The fact that the long range behavior of black holes can be well captured by minimally coupled particles can be attributed to the no hair theorem, where the classical solutions are labeled by the same set of quantum numbers as elementary particles: mass, charge and spin. Similarly, the Taub-NUT metric is described by just two quantum numbers: the mass and the NUT charge. Indeed, it was recently shown that the effective stress-energy tensor for the Kerr and the Kerr-Newman black holes are given by the classical-spin limit of minimally coupled (charged) spinning particles~\cite{Arkani-Hamed:2019ymq,Chung:2018kqs,Guevara:2018wpp,Huang}, see also~\cite{Moynihan:2019bor}. Note that while a consistent description of isolated higher spin particles beyond spin 2 is not known, the notion of minimal coupling at three points can be defined in a purely kinematic fashion~\cite{Arkani-Hamed:2017jhn}. In particular, the coupling of a spin $S$ particle to a photon or graviton is given by
\eq\label{Minimal}
\vcenter{\hbox{\includegraphics[scale=0.35]{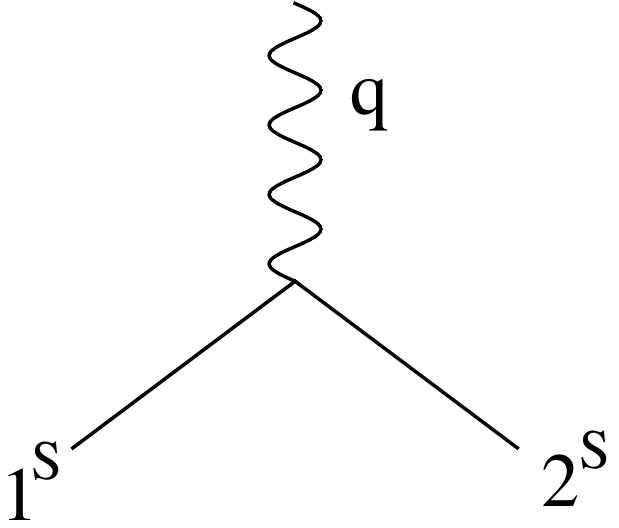}}}\quad= g(xm)^h \frac{\langle \mathbf{1}\mathbf{2}\rangle^{2S}}{m^{2S}} \,,
\eqe
where $h=(1,2)$ and $g=(\frac{\kappa}{2}, \sqrt{2}e)$, for positive photons and gravitons respectively. Here $\kappa$ is related to Newton's constant by $\kappa^2 = 32 \pi G$.

At leading order in couplings, the impulse is derived from the tree-level $2\rightarrow 2$ scattering amplitude, whose $\qb^2\rightarrow 0$ limit is captured by the product of the three-point amplitudes. Thus fundamental features of the classical solution are determined by the structure of the three-point amplitude. Indeed the correspondence between the Newman-Janis complex shift and the exponentiation of the spin in three point amplitudes is a prominent example of such a relation. Here we will consider the case where $S=0$, and instead apply a complex phase shift
\eq\label{eq:3pt}
g(xm)^h \rightarrow g(xe^{i\theta}m)^h\,.
\eqe
As mentioned in the introduction, for photon couplings, this will correspond to a eletric-magnetic duality transformation that rotates an electrically charged particle into a dyon. We will first demonstrate this by matching the impulse derived via eq.(\ref{eq:impulseGeneral}), using $h=1$ in eq.(\ref{eq:3pt}), and that from the Lorentz force sourced by the dyon solution. Next, we will take $h=2$ in eq.(\ref{eq:3pt}) instead, and show that the result matches that the impulse resulting from geodesic motion in the Taub-NUT background. This reaffirms the double copy relation between Taub-NUT and dyon solutions derived in~\cite{Luna:2015paa}. Furthermore, the fact that the Taub-NUT impulse is generated as an electric-magnetic duality transformation acting on the minimal coupling of a scalar source, is completely in accordance with the behaviour of our supertranslation map, equation~\eqref{tranPara}.

%%%%%%%%%%%%%%%%%%%%%%%%%%%%%%%%%
\subsection{The Electromagnetic Dyon Impulse}
%%%%%%%%%%%%%%%%%%%%%%%%%%%%%%%%%%%%%%%%%%%%%
%%%%%%%%%%%%%%%%%%%%%%%%%%%%%%%%%%%%%%%%%%%%%
Let us first consider the impulse a minimally coupled three point amplitude, with complex charge, imparts on a probe particle. Without loss of generality, we take probe particle $1$ to have electric charge $e_1 = n_1 e$ and no magnetic charge, which can always be achieved via a duality rotation. Thus we have the three-point coupling of particle 1 and 2 given as:
\eq
M(q^{\pm}11')=\sqrt{2} e_1 m_1 \,x^{\pm1}_{1}, \;M(q^{\pm}22')=\sqrt{2}e_2 m_2 \,x^{\pm}_2e^{\pm i \theta},
\eqe
where we identify the real and imaginary parts of $e_2 \times \exp ( i\theta )  =Q+i \Qt$ with the dyon's electric and magnetic charges $Q$ and $\Qt$, respectively.
We also take $e_2 = n_2 e $ to be an integer multiple of the fundamental charge.
This will be justified when compared to the classical computation. Note that the phase picks up an extra sign depending on the photon helicity. The $x_{1,2}$ variables denote the proportionality factor for the massless spinors. Due to momentum conservation:
\eq\label{Xdef}
(p_1+q)^2=p_1'^{2} \Rightarrow p_1\cdot q=0\; \Rightarrow \;x_1\lambda_q^{\alpha}=\frac{(p_1)^{\alpha\dot{\alpha}}\tilde{\lambda}_{q\dot{\alpha}}}{m_1} \,,
\eqe
where we have used the bi-spinor representation of the massless momentum $q^{\alpha\dot{\alpha}}=\lambda_q^{\alpha}\tilde{\lambda}_q^{\dot{\alpha}}$. Given the three-point amplitudes, their product gives the residue for the $2\rightarrow 2$ scattering in figure \ref{fig1} in the limit where $q^2\rightarrow 0$: 
\eq\label{Factor}
M_4\left(1,2{\rightarrow} 1',2'\right)|_{\footnotesize q^2{\rightarrow} 0}
=2\frac{m_1m_2 e_1 e_2}{q^2}\left(\frac{x_{1}}{x_{2}}e^{{-}i\theta}{+}\frac{x_{2}}{x_{1}}e^{i\theta}\right)
=2e^2 \frac{m_1m_2 n_1 n_2}{q^2}\left(\frac{x_{1}}{x_{2}}e^{{-}i\theta}{+}\frac{x_{2}}{x_{1}}e^{i\theta}\right)
\eqe
As we explicitly show in appendix \ref{AppBCFW}, the ratios of $x$ variables are
\begin{align}
\frac{x_{1}}{x_{2}}=  \cosh w+\frac{i\epsilon(\eta, u_1, q, u_2)}{q\cdot \eta} \,, \\
\frac{x_{2}}{x_{1}}=  \cosh w-\frac{i\epsilon(\eta, u_1, q, u_2)}{q\cdot \eta}\,,
\end{align}
where the proper velocities $u_1$ and $u_2$ are, as usual, defined by $u^\mu = p^\mu/ m$ and $w$ is the rapidity: $\cosh w=u_1\cdot u_2$. We have also defined
\[
\epsilon(a, b, c, d) \equiv \epsilon^{\mu\nu\rho\sigma} a_\mu b_\nu c_\rho d_\sigma. 
\]
Thus we find
\eqa\label{eq:dyonFourPoint}
M_4\left(1,2{\rightarrow} 1',2'\right)|_{q^2\rightarrow 0}=\frac{4 m_1m_2 e_1}{
\qb^2}\left(Q \cosh w + \Qt \frac{\epsilon(\eta, u_1, q, u_2)}{q \cdot \eta} \right) \,.
\eqae
Substituting this into eq.(\ref{eq:impulseGeneral}), we obtain the impulse
\eqa
\Delta p_1^\mu 
&=&i e_1 \int \! \dd^4\qb\; \del(\qb\cdot u_1)\del(\qb\cdot u_2)e^{{-}i\qb\cdot b}  \frac{\qb^\mu }{\qb^2}\left[Q\cosh w + \Qt \frac{\epsilon(\eta, u_1, q, u_2)}{q \cdot \eta}  \right]\,.
\eqae

A peculiarity of this expression for the physical impulse is that it apparently depends on the unphysical vector $\eta$. This phenomenon is an artifact of the particular form of the expression and we may remove $\eta$ by resolving the vector $q^\mu$ on a particular basis. To do so, let us first introduce a short-hand notation
\[
\epsilon^\mu(a, b, c) \equiv \epsilon^{\mu\nu\rho\sigma} a_\nu b_\rho c_\sigma.
\]
Then consider the basis given by $\epsilon^\mu(\eta, u_1, q)$, $\epsilon^\mu(u_1, q, u_2)$, $\epsilon^\mu(q, u_2, \eta)$, and $\epsilon^\mu(u_2, \eta, u_1)$. In view of the delta function constraint, we find that 
\begin{equation}
q^\mu = \frac{q \cdot \eta}{\epsilon(\eta, u_1, q, u_2)} \epsilon^\mu(u_1, q, u_2) + \mathcal{O}(q^2).
\end{equation}
We may neglect the $q^2$ correction as it leads to a contribution to the impulse localised at $b^\mu = 0$, outside the domain of validity of our calculation. So, we learn that the impulse is
\begin{equation}\label{Rapid}
\Delta p_1^\mu 
=i e_1 \int \! \dd^4\qb\; \del(\qb\cdot u_1)\del(\qb\cdot u_2)e^{{-}i\qb\cdot b}  \frac{1}{\qb^2}\left[Q\, \qb^\mu \, \cosh w - \Qt \, \epsilon^\mu(u_1, u_2, q)  \right]\,.
\end{equation}
It is interesting that the unphysical vector $\eta$ does not disappear from the amplitude~\eqref{eq:dyonFourPoint}. This is consistent with the observations in Caron-Huot and Zahraee's work on magnetic amplitudes~\cite{Caron-Huot:2018ape}. The observable cannot depend on such an arbitrary choice, and indeed it does not.

Now we turn to purely classical methods, with the aim of reproducing our result, equation~\eqref{Rapid}, for the impulse due to a dyon. We begin with the Lorentz force experienced by the probe particle 1
\eq
\frac{dp_1^\mu}{d\tau}=e_1  F_2^{\mu\nu}u_{1\nu},
\eqe
where once again $u_1^\nu$ is the proper velocity of the probe, and $F_2$ is the field strength sourced by the dyon. Consider the source in its rest frame so that its proper velocity is $u_2^\mu=(1,0,0,0)$. In the rest frame the electric and magnetic field of the dyon is given by
\eq\label{CoulombCharges}
F_2^{0i}=E^i=\frac{Q}{4\pi |r|^3}r^i,\quad F_2^{ij}=-\epsilon^{ijk}B_k=-\epsilon^{ijk}\frac{\Qt}{4\pi |r|^3}r_k .
\eqe
For later reference we would like to emphasize that the spatial components of the Lorentz force then take the form
\begin{equation}\label{LorentzEM}
\frac{d \bp_1}{d\tau} =
e_1  \gamma \Big(
\bE + \bv\times\bB
\Big),
\end{equation}
where $\gamma$ is the relativistic Lorentz factor.
Let $x_1=u_1\tau$ and $x_2=b{+}u_2\tau$. Upon Fourier transforming, we learn that
\eq
\widetilde{F}^{\mu\nu}_{2}=i\delta(q\cdot u_2)e^{-iq\cdot b}\frac{1}{q^2}(Qq^{[\mu}u^{\nu]}_2{-}\Qt\epsilon^{\mu\nu\rho\sigma}u_{2\rho}q_{\sigma}) \,.
\eqe
Then the impulse is simply, with $x_1(\tau)=u_1\tau$ ,  
\eqa
\Delta p_1^\mu&=&\int d\tau \frac{dp_1^\mu}{d\tau}= e_1  \int d\tau\int d^4q\, e^{-iq\cdot x} \widetilde{F}_2^{\mu\nu} u_{1\nu}\nonumber\\
&=&i e_1  \int d^4q\,\hat{\delta}(q\cdot u_1)\hat{\delta}(q\cdot u_2)\,e^{-iq\cdot  b}\frac{1}{q^2}[Qq^{\mu}(u_1\cdot u_2){-}\Qt \epsilon^{\mu}(u_{1}, u_{2},  q)]
\eqae
which indeed agrees with equation~\eqref{Rapid}. We have therefore confirmed that the phase rotation of minimal coupling with a photon gives the electromagnetic field sourced by the dyon at leading order of the electric and magnetic charge.

%%%%%%%%%%%%%%%%%%%%%%%%%%%%%%%%%%%%%%%%%%%%%%%%%%%%%%%%%%%%%%%%
\subsection{The Taub-NUT Impulse}
%%%%%%%%%%%%%%%%%%%%%%%%%%%%%%%%%%%%%%%%%%%%%%%%%%%%%%%%%%%%%%%%
%%%%%%%%%%%%%%%%%%%%%%%%%%%%%%%%%%%%%%%%%%%%%%%%%%%%%%%%%%%%%%%%

We now turn to compute the gravitational impulse induced by the corresponding phase shift. First we compute it using the formula \eqref{eq:impulseGeneral} and the three-point amplitude \eqref{Minimal}. The relevant four-point amplitude is now:
\eqa
M_4\left(1,2{\rightarrow} 1',2'\right)|_{\footnotesize q^2{\rightarrow} 0}
=\frac{\kappa ^2 }{4}\frac{m_1m_2}{q^2}\left(\frac{x^2_{1}}{x^2_{2}}e^{{-}i2\theta}{+}\frac{x^2_{2}}{x^2_{1}}e^{i2\theta}\right)=\frac{\kappa ^2 }{4}\frac{m_1m_2}{q^2}\left(e^{{-}i2\theta}e^{2w}{+}e^{i2\theta}e^{-2w}\right)\,,
\eqae
where we have expressed the ratios of $x$s in terms of rapidity to simplify the double copy, see appendix \ref{AppBCFW}. We see that the manifest double copy feature of the three-point amplitude implies that the gravitational impulse is simply given by doubling the phase $\theta$ and rapidity $w$ of the electromagnetic case, along with $\sqrt{2}e\rightarrow \frac{\kappa}{2}$.
Under the double copy structure the electric charges $e_i $ therefore map into
\begin{equation}
e_i = n_i e \qquad \longleftrightarrow \qquad M_i \equiv \frac{\kappa}{2\sqrt{2}}m_i, \qquad \qquad \text{for} \qquad  i=1,2 .
\end{equation}
We then define the gravitational analogues of the dyon's electric and magnetic charges
\begin{equation}\label{massAspects}
\begin{aligned}
Q= e_2 \cos \theta & \qquad \longleftrightarrow \qquad  Q_G = M_2 \cos 2\theta ,  \\
\Qt = e_2 \sin \theta & \qquad \longleftrightarrow \qquad  \Qt_G = M_2  \sin 2\theta.
\end{aligned}
\end{equation}
$Q_G$ and $\Qt_G$ are the mass and dual (or magnetic) mass aspects of particle 2 (the background), respectively.
The impulse is then given by\footnote{Note that there are actually two solutions for the on-shell factorization condition, corresponding to $w\leftrightarrow -w$. Due to the the $\sinh 2w$ piece, the two solutions do not give the same result. This is a reflection of the presence of Dirac string for monopole solutions. See~\cite{Caron-Huot:2018ape}.  }
\begin{align}\label{AmpG}
\Delta p_1^\mu 
&=i  M_1  \int \! d^4\qb\; \delta(\qb\cdot u_1)\delta(\qb\cdot u_2)e^{{-}i\qb\cdot b} \frac{\qb^\mu }{\qb^2}\left(Q_G\cosh 2w{-}i \Qt_G\sinh 2w \right).
\end{align}

Next we compute the impulse using the classical equation of motion of a massive test particle, namely the geodesic equation, moving in the Taub-NUT background.
We will show that it matches the result derived from the three-point amplitude \eqref{AmpG} and will provide a new perspective on the double copy structure.
A test particle of mass $M_1$ is described by its four-momentum
\begin{equation}
p^{\mu}_1= M_1 u^{\mu}_1
\end{equation}
which is given in terms of the four-velocity
\begin{equation}
u^{\mu}_1=\frac{dx^{\mu}_1 }{d\tau} = \gamma \left(1,\bv\right).
\end{equation}
The motion of the massive particle is described by the geodesic equation
\begin{equation}
\frac{d^2 x^{\mu}_1}{d\tau^2} = - \Gamma ^{\mu}_{\nu \rho} \frac{dx^{\nu}_1}{d\tau}\frac{dx^{\rho}_1}{d\tau}.
\end{equation}
The spatial components of the geodesic equation are given by
\begin{equation}\label{geoEq}
 \frac{d  \bp _1  }{d\tau} = 
- M_1   \gamma ^2\Big(  \Gamma ^{i}_{00} 
+2 \Gamma ^{i}_{0j} v^j
+ \Gamma ^{i}_{j k} v^j v^k 
\Big)
.
\end{equation}
where we have used that $\gamma = \frac{dt}{d \tau} = \left( 1-\bv\cdot \bv \right)^{-1/2}$. We now wish to evaluate the geodesic equation on the Taub-NUT background \eqref{TaubNUTMetric} in the leading post-Minkowskian approximation.
We will soon relate the standard parameterization of the Taub-NUT metric in terms of the mass $m$ and the NUT parameter $\ell$ to the mass and dual-mass aspects of the background \eqref{massAspects}, respectively.
In the post-Minkowskian approximation the metric is expanded around flat spacetime 
\begin{equation}
g_{\mu\nu} = \eta_{\mu\nu} + h_{\mu\nu},
\end{equation}
where $\eta_{\mu\nu} $ is the flat metric and at leading order the Taub-NUT background, in Cartesian coordinates, is given by
\begin{equation}\label{NPMmetric}
\begin{aligned}
h_{00}&= \frac{2m}{r}, \\
h_{0i} &=  \frac{2\ell z}{r(r^2 - z^2) } 
 \left(
\begin{tabular}{c}
$+y$   \\
$-x$    \\
$0$  
\end{tabular}
\right),
\\
h_{ij} &= \frac{2m x_i x_j}{r^3} .
\end{aligned}
\end{equation}
In polar coordinates we have $h_{0i} dx^i= -2 \ell \cos \theta d\varphi$.
The Christoffel symbols appearing in the geodesic equation \eqref{geoEq} then take the form
\begin{equation}
\begin{aligned}
\Gamma ^i_{00} &= - \frac{1}{2} \pa^i h_{00} , \\
\Gamma ^i_{0j} &= \frac{1}{2} \eta^{ik} \left(\pa_k h_{0j} - \pa_j h_{0k} \right),\\
\Gamma^i _{jk} &= \frac{2mx^i}{r^3} \delta_{jk}  - \frac{3m x^i x_j x_k}{r^5}.
\end{aligned}
\end{equation}
Notice that the first and third terms in the geodesic equation \eqref{geoEq} are proportional to the mass aspect of the background metric while the second term is proportional to its NUT charge. We will soon see that accordingly the first and third terms will contribute to the ``electric'' component of the force while the second term will contribute to the ``magnetic'' component.

We can bring the geodesic equation to a form similar to the Lorentz force in gauge theory by defining
\begin{equation}
\begin{aligned}
\phi & \equiv - \frac{1}{2}h_{00}, \\
A_i &\equiv - \frac{1}{4}h_{0i}.
\end{aligned}
\end{equation}
The potentials $\phi$ and $\bA$ are the analogues of the scalar and vector potentials in electrodynamics.
Finally, we arrive at the following expression for the geodesic equation in terms of the potentials
\begin{equation}\label{FinalGeo}
 \frac{d \bp _1}{d\tau} =
M_1 \gamma \Big(
  \frac{2\gamma ^2 -1}{\gamma} 
\bE
+4 \gamma \,  \bv\times\bB
+\bF_T
+ \bF_v
 \Big),
\end{equation}
where the electric and magnetic fields are given by
\begin{equation}\label{fields}
\begin{aligned}
\bE& =\bD \phi= -\frac{m}{r^3} \br ,\\
\bB&= \bD  \times \bA=\frac{\ell}{2 r^3} \br.
\end{aligned} 
\end{equation}
In addition to the electric and magnetic forces we also find transient and velocity forces. The transient force is given by
\begin{equation}
\left( \bF_T \right)^i = \frac{d}{d \tau} \Big( \frac{1}{2 \gamma} \eta^{ij} v^k h_{jk} \Big),
\end{equation}
which can be expressed in terms of the electric field
\begin{equation}
 \bF_T= \frac{d}{d\tau}\Big( \frac{  \left(\bv \cdot \br\right)  }{\gamma ^2} \bE \Big).
\end{equation}
The velocity force is given by
\begin{equation}
\bF_v =  - \frac{1}{\gamma} \frac{(  \bv \cdot \br )  (\bE \cdot \br ) }{  r^2 }  \bv \,,
\end{equation}
and is also proportional to the electric field. In the above derivation we have used that to leading order in the post-Minkowskian approximation the velocity is constant.

Let us now comment on the geodesic equation \eqref{FinalGeo} that we have derived.
First of all, in the post-Newtonian (PN) approximation, where velocities are small ($\gamma \rightarrow 1$), it reduces to the form of the Lorentz force in electrodynamics \eqref{LorentzEM}
\begin{equation}
\frac{d \bp _1}{d\tau} \Big| _{\text{PN}}=
M_1 \Big(
\bE +4   \bv\times\bB
\Big),
\end{equation}
where the charge of the test particle is replaced by minus its mass. Note that, in particular, both transient and the velocity forces vanish in this approximation.
In this case there is a factor of four in front of the magnetic force with respect to the electromagnetic case. This result, including the factor of four, was established long ago \cite{DeWitt:1966yi} and by now appears in many standard texts (for example \cite{Wald:1984rg}). More generally, the electric and magnetic forces appear with coefficients that depend on the Lorentz factor $\gamma$ and there are two additional forces. The transient force $\bF_T$ is a total derivative with respect to $\tau$ and therefore its contribution to the impulse vanishes since the force decays to zero at the boundaries. The velocity force $\bF_v$ is not a total derivative but its contribution to the impulse is zero as well. This can be seen by considering the transformation law of the integrated force
\begin{equation}
\int d\tau \,  \bF_v 
= -m \int  ( d \br \cdot \br ) \frac{ \bv }{\gamma^2 r^3}
\end{equation}
under the discrete CPT symmetry (charge conjugation, parity and time reversal) which takes $\br \rightarrow - \br$ and $t \rightarrow -t $. The expression above is odd under CPT and therefore vanishes.
Both forces $\bF_T$ and $\bF_v$ represent transient, short-lived, effects that do not contribute to the impulse.

Finally, we are now in a good position to compute the impulse using the geodesic equation and compare with the result derived from the three-point amplitude \eqref{AmpG}. Recall that the Lorentz factor can be expressed in terms of the rapidity
\begin{equation}
\begin{aligned}
\gamma &= \cosh w , \qquad   &  \sqrt{\gamma^2-1} = \sinh w, \\
2\gamma ^2 -1  &= \cosh 2w, \qquad   &  2\gamma \sqrt{\gamma^2-1} = \sinh 2w ,
\end{aligned}
\end{equation}
such that the geodesic equation now takes the form
\begin{equation}\label{FinalGeow}
 \frac{d \bp _1 }{d\tau} =
M_1 \gamma \Big(
\frac{\cosh 2w }{\cosh w}  
\bE
+2 \frac{\sinh 2w }{\sinh w}   \bv\times\bB
+\bF_T
+ \bF_v
\Big).
\end{equation}
Comparing the geodesic equation \eqref{FinalGeow} with the Lorentz force \eqref{LorentzEM}, we can now repeat the derivation of the impulse word by word as in the electromagnetic case, taking into account the $w$-dependent coefficients, the factor of two in front of the magnetic force and the vanishing contribution of the transient forces to the impulse.
The result is
\begin{equation}
\Delta p^{\mu} _1 = i   M_1 \int d^4 q \hat{\delta} (q\cdot u_1)\hat{\delta} (q\cdot u_2)e^{-iq \cdot b}\frac{q^{\mu}}{q^2}   \left(Q_G \cosh 2w - i \Qt_G \sinh 2w \right),
\end{equation}
where we have identified the parameters $m$ and $\ell$ with the mass and dual mass aspects \eqref{massAspects} as follows
\begin{equation}
\begin{aligned}
Q_G &= M_2 \cos 2\theta =  4\pi m ,\\
\Qt_G &=  M_2 \sin 2\theta  = 4 \pi \ell .
\end{aligned}
\end{equation}
The impulse computed using the geodesic equation agrees with the result from the amplitude's perspective \eqref{AmpG}.
Amazingly, the coefficients in front of the electric and magnetic forces in the geodesic equation \eqref{FinalGeow} precisely account for the replacement $w \rightarrow 2 w$ as anticipated from the double copy structure of the three-particle amplitude!

 %%%%%%%%%%%%%%%%%%%%%%%%%%%%%%%%%%%%%%%%%%%%%%%%%%%%%%%%%%%%%%%%
 %%%%%%%%%%%%%%%%%%%%%%%%%%%%%%%%%%%%%%%%%%%%%%%%%%%%%%%%%%%%%%%%
 \section{Discussion}\label{ConclusionSection}
 %%%%%%%%%%%%%%%%%%%%%%%%%%%%%%%%%%%%%%%%%%%%%%%%%%%%%%%%%%%%%%%%
 %%%%%%%%%%%%%%%%%%%%%%%%%%%%%%%%%%%%%%%%%%%%%%%%%%%%%%%%%%%%%%%%
In this paper, we provided a physical origin of the complex shift found by Talbot~\cite{Talbot:1969bpa}, which transforms the Schwarzschild black hole solution to the Taub-NUT metric, much in the spirit of Newman-Janis~\cite{Newman:1965tw} shift for Kerr black holes. We argued that the complex shift corresponds to an electric magnetic duality transformation, in which the Taub-NUT metric is the ``dyonic" version of the Schwarzschild solution. 
 
This was demonstrated on two fronts. First, we reinterpreted the Talbot shift as a complex BMS supertranslation. Since the transformation parameter is complex instead of real, it is not a symmetry of the asymptotic metric, but rather transforms the Schwarzschild solution to that of  Taub-NUT. Second, we showed that such complex supertranslation mixes the supertranslation charge with its dual, introduced in~\cite{Kol:2019nkc,Godazgar:2018qpq,Godazgar:2018dvh}, where the dual charge is proportional to the NUT charge. Thus the complex translation is simply an electric-magnetic duality rotation acting on the standard and dual supertranslation charges.  
 
This picture can also be nicely captured from the on-shell S-matrix point of view, where the Taub-NUT and the dyonic solution are related via the double copy in a transparent matter. In particular, we argued that their representations as three-point amplitudes between a probe and a background source, are given by a phase rotation acting on the gravitational and electromagnetic minimal coupling respectively. We confirmed this interpretation by computing the impulse imparted on a probe particle in the background of the classical sources. We proceeded by first extracting it from the $2\rightarrow 2$ scattering amplitude with single photon or graviton exchange. In view of our interest in long range effects, we took the $q^2\rightarrow0$ limit where $q$ is the momentum transfer, so that the four-point amplitude is determined by the minimally coupled three-point amplitudes. Comparing the result with that computed from the Lorentz force and geodesic equations respectively, we found an exact match.

We would like to emphasize that in this work we considered the conical singularity of the Taub-NUT metric as a physical cosmic string and studied the impulse imparted to a probe particle moving on the string's background.
We made no attempt to address the question of how to eliminate the string singularity, for example via Misner's interpretation of the Taub-NUT metric. In this way we avoid having to discuss the pathologies associated with Misner's interpretation, such as the appearance of closed timelike curves.
In this context we would like to mention the work of \cite{Kol:2020ucd}, in which an alternative to Misner's approach is presented.
It would be interesting to repeat the computation of the impulse for the background solution presented in \cite{Kol:2020ucd}, which is free of any string singularities and also devoid of closed timelike curves.

\begin{figure}[]
	\begin{center}
		\includegraphics[scale=1.25]{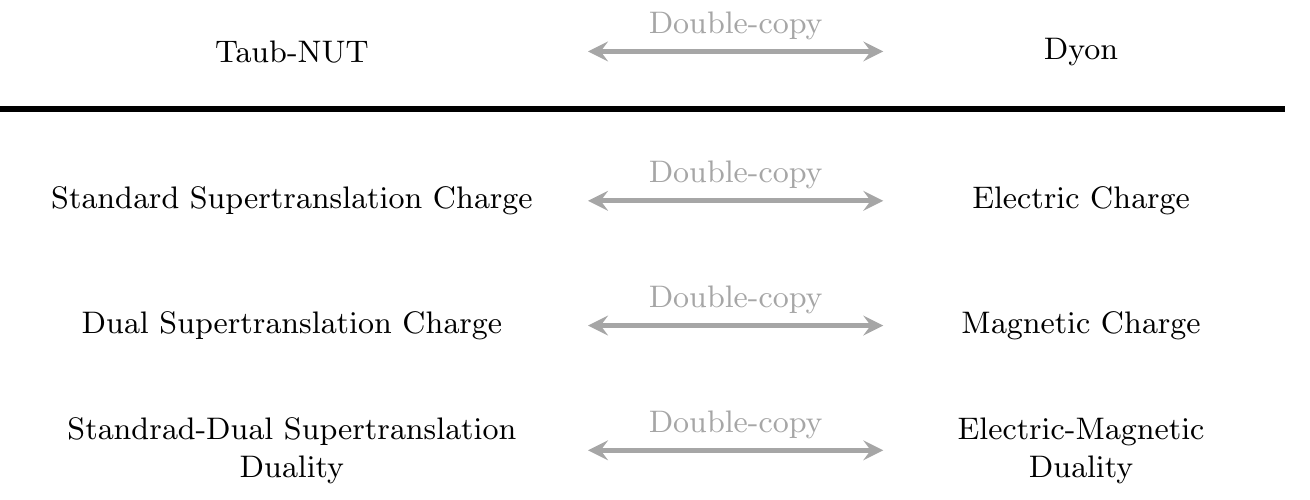}
	\end{center}
	\caption{The Taub-NUT metric admits a double copy structure which relates it to the electromagnetic dyon \cite{Luna:2015paa}. We show that the double copy structure relates the standard and dual supertranslation charges to the large electric and magnetic charges of the dyon. This result implies that the duality \eqref{duality} between the standard and dual supertranslation charges is related to the electric-magnetic duality in QED by the double copy structure.
	}
	\label{doubleCopyFig}
\end{figure}

Let us now comment on the implications of the double copy structure and how it relates charges in gauge theory to gravitational charges. In \cite{Kol:2019nkc} it was argued that the standard and dual supertranslation charges, \eqref{superT} and \eqref{dualSuperT}, are analogous to the large electric and magnetic charges in electrodynamics
  \begin{equation}
 \begin{aligned}
Q (\varepsilon) &= \frac{1}{e^2} \int \varepsilon \, *F, \\
M (\varepsilon) &= \frac{1}{2\pi} \int \varepsilon \, F,
 \end{aligned}
 \end{equation}
 respectively. We are now in a position to show that, at least in the case of the dyon and the Taub-NUT metric, the charges are not just analogous but in fact are related by the double copy. The large electric and magnetic charges of the dyon, whose fields are defined in \eqref{CoulombCharges}, are given by
  \begin{equation}
 \begin{aligned}
 Q (\varepsilon) &= \frac{1}{e^2} \frac{Q}{4\pi } \int_{S^2} d^2 z \gammaflat \, \varepsilon(z,\zb), \\
 M (\varepsilon) &= \frac{\Qt}{4\pi}  \int_{S^2} d^2 z \gammaflat \, \varepsilon(z,\zb).
 \end{aligned}
 \end{equation}
On the other hand, the standard and dual supertranslation charges of the Taub-NUT metric are given by
  \begin{equation}
\begin{aligned}
T (\varepsilon) &= \frac{m}{4\pi  G} \int_{S^2} d^2 z \gammaflat \, \varepsilon(z,\zb)= \frac{1}{4\pi  G}  \frac{Q_G}{4\pi} \int_{S^2} d^2 z \gammaflat \, \varepsilon(z,\zb), \\
\mM (\varepsilon) &= \frac{\ell}{4\pi G }  \int_{S^2} d^2 z \gammaflat \, \varepsilon(z,\zb)=  \frac{1}{4\pi  G}  \frac{\Qt_G}{4\pi}  \int_{S^2} d^2 z \gammaflat \, \varepsilon(z,\zb).
\end{aligned}
\end{equation}
Using the map between the charges \eqref{massAspects}, which is implied by the double copy structure, we can now draw a map between the large gauge charges and the large gravitational charges
   \begin{equation}
 \begin{aligned}
4 \pi G \, T(\varepsilon) &\Longleftrightarrow e ^2 Q(\varepsilon), \\
4 \pi G \, \mM(\varepsilon)  &\Longleftrightarrow M(\varepsilon).
 \end{aligned}
 \end{equation}
 This map generalizes to any metric that admits the double copy structure
 \begin{equation}\label{doubleCopyMetric}
 g_{\mu\nu}  = \eta_{\mu\nu} +2 \kappa \,  \phi \,  k_{\mu} k_{\nu},
 \end{equation}
 from which a gauge field that solves the Yang-Mills equations can be defined
 \begin{equation}
 A_{\mu} = \phi k_{\mu} .
 \end{equation}
 To see this we first write the Bondi mass and $U_z$  component in terms of the metric representation in \eqref{doubleCopyMetric}
 \begin{equation}
\begin{aligned}
 m_B & = \lim_{r \rightarrow \infty } \frac{r}{2\kappa} ( g_{uu} - \eta_{uu} ) =  \lim_{r \rightarrow \infty } r \phi k_u k_u=  \lim_{r \rightarrow \infty } r  k_u A_u , \\
 U_z &= \frac{1}{\kappa} \lim_{r \rightarrow \infty } (g_{uz}-\eta_{uz}) =  \lim_{r \rightarrow \infty } 2 \phi k_u k_z =  \lim_{r \rightarrow \infty }2  k_u A_z .
\end{aligned}
 \end{equation}
 We now use that $k_u=1$ \cite{Luna:2015paa} and re-write the standard \eqref{superT} and dual \eqref{dualSuperT}  supertranslation charges as
 \begin{equation}
\begin{aligned}
(4 \pi G) \times  T(f) &=  \int  _{\mI^+_-}  d^2 z   \gammaflat f(z,\zb) \lim_{r \rightarrow \infty } r A_u  ,\\
(4 \pi G) \times  \mM(f) &=  \int  _{\mI^+_-}  d^2 z   \gammaflat f(z,\zb) \lim_{r \rightarrow \infty }  \left(\pa_z A_{\zb}-\pa_{\zb}A_z\right) ,
\end{aligned}
 \end{equation}
which are precisely the expressions for the large electric and magnetic $U(1)$ charges, see   \cite{He:2014cra,Strominger:2015bla,Strominger:2017zoo}. The map between the large gauge charges and the large gravitational charges is described in figure \ref{doubleCopyFig}.

Finally, the analysis of this paper suggests that the more general Kerr-Taub-NUT solution should be generated from the phase rotation on the minimally coupled spinning particle, for which the classical limit of the spin induces an exponentiation effect in terms of spin operators~\cite{Arkani-Hamed:2019ymq}. An additional observable, the angular impulse (total change in spin) during scattering, is available in this more general case, providing more sensitivity to the detailed structure of the metric. This angular impulse, like the usual linear impulse we discussed above, can be computed from scattering amplitudes~\cite{Maybee:2019jus,Guevara:2019fsj}. It would be interesting to verify this conjectural link to Kerr-Taub-NUT by comparing the classical linear and angular impulses with those computed via geodesic equations, utilizing the simplifications in the post Minkowskian expansion discussed in this paper. As a further step, one could consider solutions of the combined Einstein-Maxwell equations which again connect to minimally coupled scattering amplitudes~\cite{Moynihan:2019bor}. Indeed considerations from the Weyl spinor formulation of the classical double copy~\cite{Luna:2018dpt} suggest that similar ideas apply to the full Plebanski-Demianski~\cite{Debever:1971,Plebanski:1976gy} class of solutions, which include not only spin, electric and magnetic charges, masses and NUT parameters, but also an acceleration parameter as well as a cosmological constant.

 \section*{Acknowledgements}

 UK would like to thank Massimo Porrati for a collaboration on related subjects. DOC thanks Tim Adamo for useful comments, and a great deal of education.
 YTH is supported by MoST Grant No. 106-2628-M-002-012-MY3. YTH is also supported by Golden Jade fellowship.
 The research of UK is supported in part by NSF through grant PHY-1915219.
 Meanwhile DOC is supported by the STFC grant ``Particle Theory at the Higgs Centre''.

 \appendix

 \section{Complex Coordinates Transformations}\label{appSec}

Newman and Janis \cite{Newman:1965tw} found that the Kerr solution of a spinning black hole can be derived from the Schwarzschild solution using a complex coordinate transformation. Talbot \cite{Talbot:1969bpa} had generalized their results and in particular he found a complex coordinates transformation that maps Schwarzschild into Taub-NUT. We start by review the general algorithm for the map between different solutions (for recent reviews and works on the subject see \cite{Adamo:2014baa}-\cite{Erbin:2016lzq}).

 The algorithm is based on the Newman-Penrose (NP) formalism, which is defined using a set of complex null tetrads
 \begin{equation}
 Z^{\mu}_a = \{ \ell^{\mu},n^{\mu},m^{\mu},\mb^{\mu}\},
 \end{equation}
 where $\mb^{\mu} = (m^{\mu })^*$.
 Note that while the tetrads are complex in the NP formalism, the metric is always real and at this point the spacetime coordinates are still real as well.
 The inverse metric is given by
 \begin{equation}
 g^{\mu\nu} = \eta^{ab} Z^{\mu}_a Z^{\nu}_b
 \end{equation}
 with
 \begin{equation}
 \eta^{ab } =\left(
 \begin{matrix} 
 0 & -1 & 0 & 0 \\
 -1 & 0 & 0 & 0 \\
 0 & 0 & 0 & 1 \\
 0 & 0 & 1 & 0 
 \end{matrix}
 \right),
 \end{equation}
 such that
 \begin{equation}
 g^{\mu\nu} = - \ell^{\mu} n^{\nu}- \ell^{\nu} n^{\mu} + m^{\mu} \mb^{\nu}+ m^{\nu} \mb^{\mu}.
 \end{equation}

 The map is defined using the following algorithm:
 \subsection*{1. The seed metric}
 The seed metric is given by
 \begin{equation}
 ds^2 = -f(r) du^2 -2 du dr +r^2 d\Omega^2,
 \end{equation}
 where $u$ is a null coordinate and $r$ is a radial coordinate. The null tetrads of the metric are
 \begin{equation}
 \ell^{\mu} = \delta^{\mu}_r , \qquad
 n^{\mu}= \delta^{\mu}_u - \frac{f}{2} \delta^{\mu}_r ,\qquad
 m^{\mu} = \frac{1}{\sqrt{2} r} \left( \delta^{\mu}_{\theta} + \frac{i}{\sin \theta} \delta^{\mu}_{\phi} \right).
 \end{equation}
 We will take the seed metric to be Schwarzschild, namely
 \begin{equation}
 f(r) = 1- \frac{2m}{r}.
 \end{equation}

 \subsection*{2. Complexification of two coordinates}
 We now allow the coordinates $u$ and $r$ to take complex values while insisting that the metric remains real (namely $\ell^{\mu},n^{\mu}$ must still be real and $m^{\mu}$ and $\mb^{\mu}$ must still be complex conjugates of each other). In addition, the function $f(r)$ is transformed to
 \begin{equation}
 f(r) \rightarrow f(r, \rb) = 1- \frac{m}{r} - \frac{\mb}{\rb}.
 \end{equation}
 There are many ways to complexify the wrap function. The choice above proved to produce good solutions but generally speaking this step is arbitrary and it is not known if there is a rigorous way to prove it.

 \subsection*{3. Complex coordinates transformation}
 Once we allowed the coordinates to take complex values we can perform a complex coordinate transformation
 \begin{equation}\label{coTrans}
 u= u'  - i a \cos \theta + 2 i \ell \log \sin \theta, \qquad
 r = r' + i a \cos \theta - i \ell ,
 \end{equation}
 accompanied by the following map of the mass parameter
 \begin{equation}
 m=m' - i \ell.
 \end{equation}
 Under this transformation the wrap function takes the form
 \begin{equation}
 f= 1- \frac{2mr' +2 \ell (\ell -a \cos \theta)}{\rho^2}, \qquad \rho ^2 = r' \,\! ^2 +(\ell -a \cos \theta)^2,
 \end{equation}
 and the tetrads, which transform as vectors
 \begin{equation}
 Z_a^{'\mu} = \frac{\pa x^{'\mu}}{\pa x^{\nu}} Z^{\nu}_a,
 \end{equation}
 are now given by
 \begin{equation}
 \begin{aligned}
 \ell^{\mu} &= \delta^{\mu}_r , \qquad
 n^{\mu}= \delta^{\mu}_{u'} - \frac{f}{2} \delta^{\mu}_{r'} , \\
 m^{\mu} &= \frac{1}{\sqrt{2} (r'+i a \cos \theta -i \ell)} \left( \delta^{\mu}_{\theta} + \frac{i}{\sin \theta} \delta^{\mu}_{\phi} 
 +i (a \sin \theta +2 \ell \cot \theta) \delta ^{\mu}_{u'}
 -i a \sin \theta \delta^{\mu}_{r'}
 \right).
 \end{aligned}
 \end{equation}
 We now impose that the coordinates $u'$ and $r'$ take real values.

 \subsection*{4. Reconstructing the metric}
 Having the new tetrad basis at hand we can reconstruct the inverse metric and then invert it to get
 \begin{equation}
 \begin{aligned}
 ds^2 = & -f \left(du' + \Omega d \phi + \frac{r' \, \! ^2 +a^2 +\ell^2 +a \Omega}{\Delta}dr' \right)^2
 +\frac{\rho^2}{\Delta} dr' \,\! ^2 \\
 &+\rho^2 \left(d\theta ^2 + \sigma ^2 \sin ^2 \theta \left(d\phi + \frac{a}{\Delta} dr' \right) ^2 \right) ^2 ,
 \end{aligned}
 \end{equation}
 where
 \begin{equation}
 \Omega = -2 \ell \cos \theta - (1-f^{-1}) a \sin ^2 \theta, \qquad    \Delta = r' \,\! ^2 -2mr' +a ^2 -\ell^2, \qquad    \sigma^2 =\frac{\Delta}{f \rho^2}.
 \end{equation}

 \subsection*{5. Boyer-Lindquist coordinates}
 Finally, using the following change of coordinates
 \begin{equation}
 \begin{aligned}
 u' &\rightarrow u' - m \log \left(r'\,\! ^2-2mr'-l^2 +a^2\right) - 2\frac{m^2+\ell^2}{\sqrt{a^2 -m^2 -\ell^2}} \arctan \frac{r'-m}{\sqrt{a^2 -m^2 -\ell^2}} ,\\
 \phi& \rightarrow \phi - \frac{a}{\sqrt{a^2 -m^2 -\ell^2}} \arctan \frac{r'-m}{\sqrt{a^2 -m^2 -\ell^2}},
 \end{aligned}
 \end{equation}
 one can transform the metric into the more standard Boyer-Lindquist form
 \begin{equation}\label{TNKErr}
 ds^2 =  -f \left(du' +dr' + \Omega d \phi  \right)^2
 +\frac{\rho^2}{\Delta} dr' \,\! ^2 +\rho^2 \left(d\theta ^2 + \sigma ^2 \sin ^2 \theta d\phi  ^2 \right) ^2 .
 \end{equation}
 The metric \eqref{TNKErr} describes the  Taub-NUT-Kerr solution.
 
 Let us stress that the use of the tetard formalism is essential for the algorithm to work. The complex change of coordinates is not applied directly on the metric but rather on the tetrads.
%%%%%%%%%%%%%%%%%%%%%%%%%%%%%%%%%%%%%%%%
\section{$\frac{x_1}{x_2}$ and rapidity}\label{AppBCFW}
%%%%%%%%%%%%%%%%%%%%%%%%%%%%%%%%%%%%%%%%
 In this appendix we derive the express the little group invariant the ratio of $\frac{x_1}{x_2}$ in terms of momentum factors. First, from the definition of $x$ in eq.(\ref{Xdef}) we have
 \eq
 x_1=\frac{\langle \eta|p_1|q]}{m_1\langle \eta q\rangle},\quad \frac{1}{x_2}=\frac{[\tilde\eta|p_2|q\rangle}{m_2[\tilde\eta q]}\,.
 \eqe 
where $|\eta\rangle,|\tilde\eta]$ are some auxiliary spinors such that $\langle \eta q\rangle$ and $[\tilde\eta q]\neq0$, and we have $\eta^{\alpha\dot{\alpha}}=|\eta\rangle[\tilde\eta|$. Putting the two together, we have:
\eqa
\frac{x_1}{x_2}&=&-\frac{1}{m_1m_2}\frac{\langle \eta|p_1|q]}{\langle \eta q\rangle}\frac{\langle q|p_2|\eta]}{[q\eta]}=-\frac{1}{m_1m_2}\frac{Tr\left[(1-\gamma_5) \eta p_1 q p_2\right]}{4(q\cdot \eta)}\nonumber\\
&=&-\frac{(\eta \cdot p_1)(q\cdot p_2)-(\eta \cdot q)(p_2\cdot p_1)+(\eta \cdot p_2)(p_1\cdot q)-i\epsilon(\eta p_1 q p_2)}{m_1m_2(q\cdot \eta)}\nonumber\\
&=&\frac{(p_2\cdot p_1)}{m_1m_2}+\frac{i\epsilon(\eta p_1 q p_2)}{m_1m_2(q\cdot \eta)}
\eqae 
where coming to the last line we used the fact that $(q\cdot p_2)=(q\cdot p_1)=0$. 

Finally, it is useful to convert the above in terms of rapidity for the purpose of double copy. Using 
\eq
\epsilon(\eta p_1 q p_2)^2=m^2_1m^2_2(q\cdot \eta)^2\left(1-\frac{(p_2\cdot p_1)^2}{m^2_1m^2_2}\right)
\eqe 
we see that 
\eq
\frac{x_1}{x_2}=\frac{(p_2\cdot p_1)}{m_1m_2}\pm\sqrt{\left(\frac{(p_2\cdot p_1)}{m_1m_2}\right)^2-1}=\cosh w\pm\sinh w=e^{\pm w}
\eqe
where $\cosh w=\frac{(p_2\cdot p_1)}{m_1m_2}$. Note that in terms of rapidity, there are two solutions. As discussed in the text for magnetic charges, the amplitude will have non-trivial dependence on this sign, reflecting the presence of the Dirac string. 
%%%%%%%%%%%%%%%%%%%%%%%%%%%%%%%%%%%%%%%%%%%%%%%%%%%

\end{document}